\documentclass[preprint,11pt,3p]{elsarticle}

\usepackage{amssymb}

\usepackage{lineno}

\journal{NeuroImage}

\usepackage{url,hyperref,microtype,subcaption}
\usepackage[onehalfspacing]{setspace}
\usepackage[nolist]{acronym}
\usepackage{siunitx}
\usepackage{caption}
\usepackage{romannum}
\usepackage{csquotes}
\usepackage{mathtools}
\usepackage{tabularx}
\usepackage{subcaption}
\usepackage{xcolor}
\usepackage{soul}
\usepackage{svg}
\usepackage{textcomp}
\usepackage{sidecap}
\usepackage{numprint}
\usepackage{sidecap}

\usepackage{xspace}

\newcommand{\sref}[1]{Sec.~\ref{#1}}
\newcommand{\fref}[1]{Fig.~\ref{#1}}

\newcommand{\tref}[1]{Table \ref{#1}}

\newcommand{\mm}{\si{\milli\metre}}
\newcommand{\micron}{\si{\micro\metre}}
\newcommand{\cm}{\si{\centi\metre}}

\newcommand{\area}[1]{\ensuremath{#1}}
\newcommand{\apl}{\area{a}\xspace}
\newcommand{\bg}{\area{bg}\xspace}
\newcommand{\cor}{\area{cor}\xspace}
\newcommand{\wm}{\area{wm}\xspace}
\newcommand{\vone}{\area{hOc1}\xspace}
\newcommand{\vtwo}{\area{hOc2}\xspace}
\newcommand{\vthree}{\area{hOc3v}\xspace}
\newcommand{\vfive}{\area{hOc5}\xspace}
\newcommand{\opfive}{\area{Op5}\xspace}
\newcommand{\opsix}{\area{Op6}\xspace}
\newcommand{\opseven}{\area{Op7}\xspace}
\newcommand{\areaffour}{\area{44}\xspace}
\newcommand{\areaffive}{\area{45}\xspace}

\newcommand{\hipfive}{\area{hIP5}\xspace}
\newcommand{\hipsix}{\area{hIP6}\xspace}
\newcommand{\hipseven}{\area{hIP7}\xspace}
\newcommand{\hipeight}{\area{hIP8}\xspace}
\newcommand{\sixdone}{\area{6d1}\xspace}
\newcommand{\sixdtwo}{\area{6d2}\xspace}
\newcommand{\sixdthree}{\area{6d3}\xspace}
\newcommand{\sma}{\area{SMA}\xspace}
\newcommand{\presma}{\area{preSMA}\xspace}

\newcommand{\bone}{\textsc{\itshape B01}\xspace}
\newcommand{\bb}{\textsc{\itshape B20}\xspace}
\newcommand{\ba}{\textsc{\itshape AAHB}\xspace}

\newcommand{\set}[1]{\ensuremath{\left\{#1\right\}}}

\newcommand{\interv}[2]{\ensuremath{\left[#1, #2\right]}}
\newcommand{\model}[3]{\ensuremath{f_{\interv{#1}{#2}}^{#3}}}

\newcommand{\hr}{\textsc{\itshape HR}\xspace}
\newcommand{\hrall}{\textsc{\itshape HR (all)}\xspace}
\newcommand{\lr}{\textsc{\itshape LR}\xspace}
\newcommand{\ms}{\textsc{\itshape MS}\xspace}
\newcommand{\hre}{\ensuremath{E_{HR}}\xspace}
\newcommand{\lre}{\ensuremath{E_{LR}}\xspace}

\graphicspath{{images/}}

\begin{document}
\begin{acronym}[ReLU]
    \acro{HRE}[HRE]{High Resolution Encoder}
    \acro{LRE}[LRE]{Low Resolution Encoder}
    \acro{CNN}{Convolutional Neural Network}
    \acroplural{CNN}[CNNs]{Convolutional Neural Networks}
    \acro{ReLU}[ReLU]{Rectified Linear Unit}
    \acro{ROI}[ROI]{region of interest}
    \acroplural{ROI}[ROIs]{regions of interest}
    \acro{RANSAC}[RANSAC]{Random Sample Consensus}
    \acro{FCN}[FCN]{Fully Convolutional Network}
    \acroplural{FCN}[FCNs]{Full Convolutional Networks}
    \acro{MPI}[MPI]{Message Passing Interface}
    \acro{MRI}[MRI]{Magnetic Ressonance Imaging}
\end{acronym}

\begin{frontmatter}



\title{Convolutional Neural Networks for cytoarchitectonic brain mapping at large scale}


\author[FZJ,AI]{Christian~Schiffer\corref{COR}}
\ead{c.schiffer@fz-juelich.de}

\author[ICB]{Hannah~Spitzer}
\author[CO]{Kai~Kiwitz}
\author[CO]{Nina~Unger}
\author[LO]{Konrad~Wagstyl}
\author[MNI]{Alan~C.~Evans}
\author[HHU]{Stefan~Harmeling}
\author[FZJ,CO]{Katrin~Amunts}
\author[FZJ,AI]{Timo~Dickscheid}

\cortext[COR]{Corresponding author. Institute of Neuroscience and Medicine (INM-1), Research Centre Jülich, Jülich, Germany.}
\address[FZJ]{Institute of Neuroscience and Medicine (INM-1), Research Centre Jülich, Germany}
\address[AI]{Helmholtz AI, Research Centre Jülich, Germany}
\address[ICB]{Institute of Computational Biology, Helmholtz Zentrum München, Germany}
\address[CO]{C\'{e}cile \& Oscar Vogt Institute for Brain Research, University Hospital Düsseldorf, Heinrich-Heine-University Düsseldorf, Germany}
\address[LO]{Wellcome Centre for Human Neuroimaging, University College London, London, United Kingdom}
\address[MNI]{Department of Neurology \& Neurosurgery, Montréal Neurological Institute (MNI), McGill University, Montréal, Canada}
\address[HHU]{Institute of Computer Science, Heinrich-Heine-University Düsseldorf, Germany}

\begin{abstract}
Human brain atlases provide spatial reference systems for data characterizing brain organization at different levels, coming from different brains.
Cytoarchitecture is a basic principle of the microstructural organization of the brain, as regional differences in the arrangement and composition of neuronal cells are indicators of changes in connectivity and function.
Automated scanning procedures and observer-independent methods are prerequisites to reliably identify cytoarchitectonic areas, and to achieve reproducible models of brain segregation.
Time becomes a key factor when moving from the analysis of single regions of interest towards high-throughput scanning of large series of whole-brain sections.
Here we present a new workflow for mapping cytoarchitectonic areas in large series of cell-body stained histological sections of human postmortem brains.
It is based on a Deep Convolutional Neural Network (CNN), which is trained on a pair of section images with annotations, with a large number of un-annotated sections in between.
The model learns to create all missing annotations in between with high accuracy, and faster than our previous workflow based on observer-independent mapping.
The new workflow does not require preceding 3D-reconstruction of sections, and is robust against histological artefacts.
It processes large data sets with sizes in the order of multiple Terabytes efficiently.
The workflow was integrated into a web interface, to allow access without expertise in deep learning and batch computing.
Applying deep neural networks for cytoarchitectonic mapping opens new perspectives to enable high-resolution models of brain areas, introducing CNNs to identify borders of brain areas.

\end{abstract}




\end{frontmatter}



\section{Introduction}%
\label{sec:introduction}

Human brain atlases provide a spatial framework for localizing information retrieved from neuroscientific studies of different brains, addressing brain organization from different angles and including different data modalities.
The cerebral cortex of the brain is organized into cortical areas, which each have a specific functional role.
They can be identified in cell body stained sections based on cytoarchitecture.
Regional differences in the spatial arrangement and composition of the cells covary with changes in connectivity and function~\cite[]{Goulas2018}.
Cytoarchitectonic borders can be identified in microscopic scans of histological brain sections, based on the analysis of the arrangement and distribution of cells, their different morphology and size, as well as differences in the appearance and relative thickness of cortical layers.
Such criteria have been formulated for the first time more than a century ago to map the cerebral cortex, and still serve as guidelines for cytoarchitectonic analysis~\cite[]{Amunts2015} .
Different approaches have been proposed in the past to identify positions of borders in a reliable manner~\cite[]{Schleicher1999, Schmitt2002, Annese2004}.
The de-facto standard for identifying borders of cytoarchitectonic areas in the human cerebral cortex is a method based on multivariate statistical image analysis~\cite[]{Schleicher1999}, which has been applied for the identification of more than 200 areas to date~\cite[]{Amunts2020}.
To map the whole extent of an area in both hemispheres, and to capture its intersubject variability through studies in large samples, however, is extremely time- and labor-intensive:
Cytoarchitectonic maps need to aggregate properties across many histological sections and multiple brains.
To address this challenge, mapping includes a subset of histological sections (every 15-60ths section, i.e. $0.3\mm$ to $1.2\mm$ distance between sections) of ten human postmortem brains resulting in analyses of several hundred sections per area, which corresponds to a workload in the order of one or even several person years per area~\cite[]{Amunts2020}.

Recent high-throughput scanning devices and powerful compute resources enable a much higher degree of automation in digitalization and analysis of whole human brain sections at microscopical resolution.
Technological progress has made it possible to 3D-reconstruct a complete postmortem brain at 20 micron spatial resolution with more than 7000 sections - the BigBrain~\cite[]{Amunts2013}.
This high-resolution brain model opens the possibility to produce complete maps of cytoarchitectonic areas at full microscopic resolution, and to cover large image stacks with brain areas extending across thousands of sections
Hereby, each section image has up to $\numprint{120000} \times \numprint{80000}$ pixels image size each.
In order to address these challenges, a method is required, which
\begin{enumerate}
	\item automatically classifies brain areas based on cytoarchitectonic criteria,
	\item handles series with thousands of 2D images of histological sections with data in the Giga- to Terabyte range,
	\item is robust against histological artefacts, which are inevitable in large section series,
	\item provides stable results independently of the cutting plane, e.g. when changes in the cutting direction relative to the brain tissue prevents analysis of the 6-layered structure of the cerebral cortex (in the following referred to as \emph{oblique cuts}), and
	\item can be operated and supervised by neuroscience experts without requiring advanced computer science skills.
\end{enumerate}

Previous experience in cytoarchitectonic mapping has shown that the identification of brain areas considers multiple parameters.
This is true for traditional visual inspection using a light microscope, as well as for automated mapping approaches.
It involves complex multi-scale texture patterns, from the level of neurons up to a level of cortical layers and areas.
However, several parameters that can be used for identification of cortical areas heavily depend on the cutting plane of the histological sections with respect to the orientation of cortical columns.
The highly folded cerebral cortex of the human brain hereby poses particular challenges, since brain areas may appear in a very different way in dependence on the cutting angle.
Thus, brain mapping needs to operate in a variable data space, where no restrictions should be made on the orientation of the cutting plane relative to the course of cortical layers and the brain surface.
In addition, automated brain mapping needs to consider variation in tissue quality and staining, as well as histological artefacts.
Finally, automated mapping methods must take into account variations in cytoarchitecture between different brains and lead to identical parcellations, even if interindividual differences in cytoarchitecture are large.

Previous work on automated cytoarchitectonic area segmentation~\cite{Spitzer2017, Spitzer2018} proposes to use \acp{CNN} for automatic segmentation of multiple cytoarchitectonic areas across multiple human brains.
This is a remarkably challenging task, as the model needs to be robust against the considerable interindividual variability of the human brain, inevitable histological artefacts, variations in staining, and oblique cuts, to name only a few of the constraints.
At the same time, it has to be highly sensitive to variations of cytoarchitecture in different brain areas, which may be subtle.
This may result in a need for large amounts of training data, which is difficult to cover.
Consequently, such generalized segmentation models are still subject to active research.

\begin{figure*}
	\centering
	\includegraphics[width=1.0\linewidth]{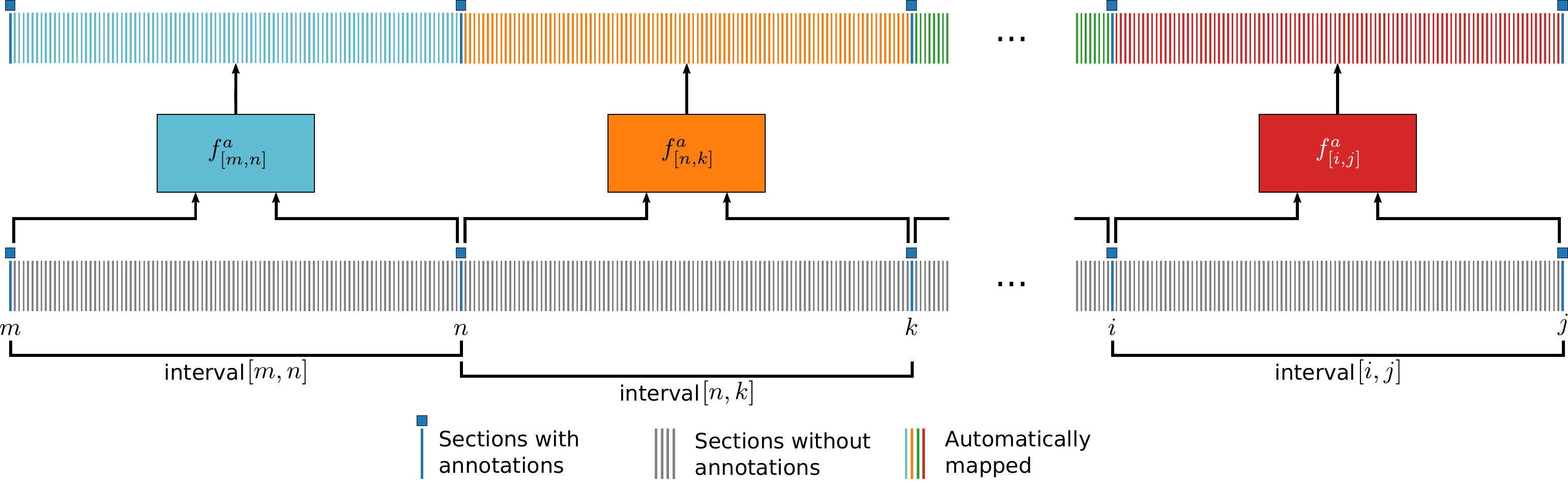}
	\caption{Setup of our workflow.
		Images of histological sections are depicted as thin vertical bars, neural network models are depicted as colored boxes.
		The full extent of sections containing a target brain area \apl (sections $m$ to $j$, bottom row) is subdivided into section intervals, which are defined by annotations at regular intervals (blue squares, $m, n, k, \ldots$).
		One local segmentation model \model{m}{n}{\apl} is trained for each interval enclosed by a pair of annotations $\interv{m}{n}$.
	After training, each model is applied to automatically map sections falling into the corresponding interval.}
	\label{fig:intervals}
\end{figure*}

We here propose a new workflow for cytoarchitectonic mapping of a \emph{target area} across large or complete series of histological human brain sections with a high degree of automation.
The workflow is illustrated in \fref{fig:intervals}.
Following a \enquote{divide \& conquer} approach, the full extent of a target brain area \apl is subdivided into intervals of sections, which are enclosed by annotations created at approximately regular section intervals.
Separate \acp{CNN} are then trained for each interval, using the enclosing annotations as training data.
This results in a set of \emph{local segmentation models}, each specialized to automatically map only the tissue sections which fall into the corresponding interval.
By training local models for each interval of target area $\apl$, an interactive workflow is obtained that allows an expert to label cytoarchitectonic areas in full stacks of histological sections with minimal manual annotation, aided by Deep Learning, and at a speed that matches high throughput image acquisition.

In this work, we
\begin{enumerate}
	\item introduce a method to automatically map cytoarchitectonic brain areas across large series of histological human brain sections (\sref{sec:materials_and_methods}),
	\item evaluate its precision on 18 cytoarchitectonic areas from the BigBrain dataset~\cite[]{Amunts2013} to investigate its applicability to a wide range of different brain areas,
	\item assess its precision for two areas in three brains with variable staining protocols~\cite[]{Amunts2013, Amunts2000, Ding2016} to investigate robustness against interindividual differences and different staining procedures, and
	\item create highly detailed and complete 3D maps of four areas in the BigBrain dataset and evaluate their anatomical plausibility\footnote{The maps are released in the public domain as part of the multilevel human brain atlas in the EBRAINS platform\url{https://www.ebrains.eu}}.
\end{enumerate}

\section{Materials and methods}%
\label{sec:materials_and_methods}

\subsection{GLI-based mapping of cytoarchitectonic areas for
training and validation}%
\label{sub:observer_independent_mapping_according_to_gli_profiles}

Our proposed method requires annotations of the target area at roughly regular intervals in approximately 1\% of sections in the stack.
Such annotations consist of localizations of areal borders in the section, and are defined using the well-established GLI-based mapping procedure described in~\cite[]{Schleicher1999}.
This approach starts by scanning the histological images and by building a Gray Level Index (GLI) image~\cite[]{Schleicher1999}.
The GLI is a measure of the volume fraction of cell bodies~\cite[]{Wree1982}.
In a next step, profiles extending from the cortical surface to the white matter border are extracted along Laplacians, which reflect laminar changes in the volume fraction of cell bodies, and thus encode cytoarchitecture.
The cortical surface and the white matter border are manually identified.
Using a sliding window procedure across the cortical ribbon, the similarity of blocks of profiles is being estimated by the Mahalanobis distance, a multi-variate distance measure, at each position, that is combined with a Hotelling's t-test for checking significance.
Borders between areas are indicated by significant peaks in the Mahalanobis distance function.
The positions of borders are then labeled in the image.
These borders are then used as a basis for the network training and validation.

\subsection{Datasets}%
\label{sub:datasets}

\begin{figure}
	\centering
	\includegraphics[width=\linewidth]{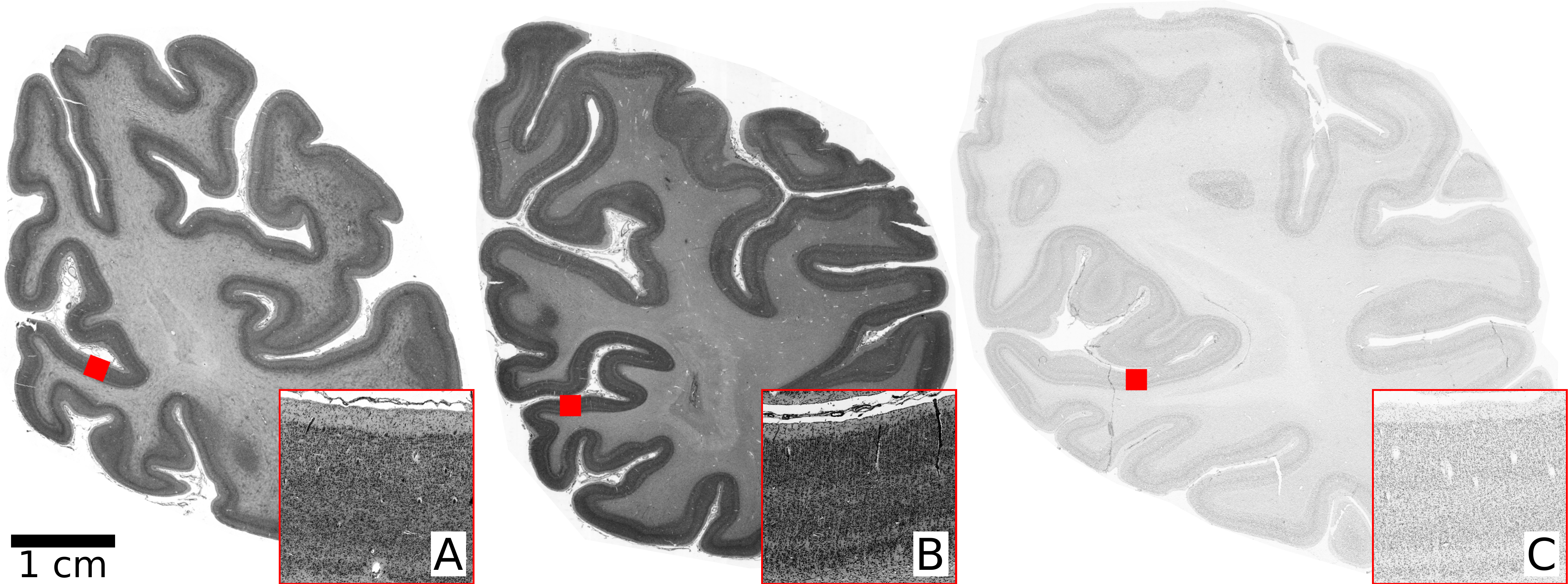}
	\caption{Example images of cell body stained histological human brain sections taken from datasets \bb (A), \bone (B) and \ba{} (C).
		 All sections were sampled from a comparable region of the occipital lobe.
		 Differences arise from intersubject variability and variations in staining and histological processing protocols.
		 Locations of detail views ($2\mm \times 2\mm$) are marked with red squares.
		 For \bb and \bone, only the right hemisphere is shown.
		 \ba{} only includes a single hemisphere.
		 Cerebellum was removed from \bb and \ba{} for visualization.
		 Scale bar: $1\cm$ (same for all three sections).
	}
	\label{fig:bb_b01_samples}
\end{figure}

The datasets used in this study comprise image series of histological sections of three human brains, which have been stained for neuronal cell bodies~\cite[]{Amunts2020, Ding2016}.
The brains vary in terms of cytoarchitecture and folding pattern, as well as staining properties, presence of histological artifacts and other features (\fref{fig:bb_b01_samples}).
Areas have been mapped in the past (cf.~Sec.~\ref{sub:observer_independent_mapping_according_to_gli_profiles}) using at least every 60th section of the series.
These maps provide the basis to train the neural network models and to perform automatic segmentation in previously unseen, close by sections.

The first dataset - denoted as \bb - is based on the original histological sections of the publicly available microscopic 3D model BigBrain~\cite{Amunts2013}{}.
The dataset consists of images of 7404 coronal sections with a thickness of $20\micron$.
A modified Merker stain~\cite[]{Merker1983} was used to stain cell bodies.
A subset of sections was scanned at $1\micron$ resolution using a high-throughput light-microscopic scanner (TissueScope HS, Huron Digital Pathology Inc.).
Annotations based on the GLI-based method (\sref{sub:observer_independent_mapping_according_to_gli_profiles}) at an interval of approximately 60 sections ($\approx 1.2\mm$) were obtained for 18 cortical areas, belonging to different functional systems:
\begin{enumerate}
	\item Visual areas \vone{}, \vtwo{}~\cite{Amunts2000}{}, \vthree{}~\cite{Rottschy2007}{} and \vfive{}~\cite{Malikovic2007}{}. Additional annotations at an interval of approximately $30$ ($0.6\mm$) sections were created for \vfive{}, as well as on a small set of sections containing \vthree{}~\cite{Kiwitza,Kiwitzb,Kiwitzc,Kiwitzd}.
	\item Areas of the frontal operculum \opfive{}, \opsix{} and \opseven{}~\cite[]{Ungera, Ungerb, Ungerc}.
	\item Areas \areaffour{} and \areaffive{} of Broca's region~\cite{Amunts1999, Amunts2004}{} in the inferior frontal gyrus.
	\item Areas \hipfive{}, \hipsix{}, \hipseven{} and \hipeight{}~\cite{Richter2019}{} in the intraparietal sulcus.
	\item Supplementary motor area \sma{} and pre-supplementary motor area \presma{}~\cite[]{Ruan2018}.
	\item Premotor areas \sixdone{}, \sixdtwo{} and \sixdthree{}~\cite[]{Sigl2018, Sigl, Sigla, Siglb} .
\end{enumerate}
The BigBrain dataset has been fully reconstructed at $20\micron$~\cite[]{Amunts2013} and therefore opens the possibility to investigate the 3D consistency of the computed maps after transformation into the reconstructed space.

Brain areas differ in cytoarchitecture, as well as in size and in how much the morphology of an area changes across a series of consecutive brain sections.
This has implications for the amount of annotations required to capture the relevant properties of certain areas.
For example, \vone is large and shows only moderate changes across consecutive sections.
In comparison, \vfive is considerably smaller, and \vthree changes considerably across consecutive sections (see \fref{fig:challenges}, C-F), resulting in a need for more annotations to capture their structure.

The second dataset - \bone - has also been used for mapping in the past, whereby every 15th section of the whole series of sections was stained and digitized.
This brain was 3D reconstructed with a spatial resolution of $1\mm$ isotropic~\cite[]{Amunts2020}.
Annotations for visual areas \vone{} and \vtwo{} at an interval of approximately every 60th section~\cite{Amunts2000}{} in a subset of sections have been used.
This dataset serves to investigate robustness against intersubject variability, while the lab protocol is similar to the one used for \bb.

The third dataset - \ba -, comes from the \emph{Allen Adult Human Brain Atlas}~\cite[]{Ding2016}.
It includes 106 unevenly spaced, publicly available sections.
In contrast to the first two series of images, it differs in thickness ($50\micron$), and the staining method (Nissl staining).
Annotations are provided for cortical and subcortical gray matter according to a modified Brodmann scheme on one hemisphere (cf.~\cite{Ding2016}{}).
This dataset is used to investigate robustness of the proposed method against variable lab protocols and delineation criteria with respect to areas \vone{} and \vtwo{}, which correspond to \textit{\enquote{primary visual cortex (striate cortex, area V1/17)}} (identifier 10269) and \textit{\enquote{parastriate cortex (area V2, area 18)}} (identifier 10271), respectively, in the Allen ontology.

\subsection{Local segmentation models}%
\label{sub:local_segmentation_models}

Annotations of cytoarchitectonic areas based on GLI mapping (\sref{sub:observer_independent_mapping_according_to_gli_profiles}) were used to train \acp{CNN}, which we refer to as \emph{local segmentation models}.
Each local segmentation model \model{s_1}{s_2}{\apl} was trained on two sections $s_1$ and $s_2$ (the \emph{training sections}) with available annotations for a target area \apl.
Trained local segmentation models were then applied to \enquote{fill the gaps}, i.e. to automatically segment the target area in sections enclosed by the respective training sections $s_1$ and $s_2$ (\fref{fig:intervals}).
The focus on a single target area and a spatially restricted stack of consecutive sections reduces cytoarchitectonic and morphological variations that need to be captured by the respective models, which we expect to result in improved performance compared to training models for multiple areas or a wider range of sections as proposed in \cite[]{Spitzer2017}.

\begin{figure*}[t]
	\centering
	\includegraphics[width=0.95\linewidth]{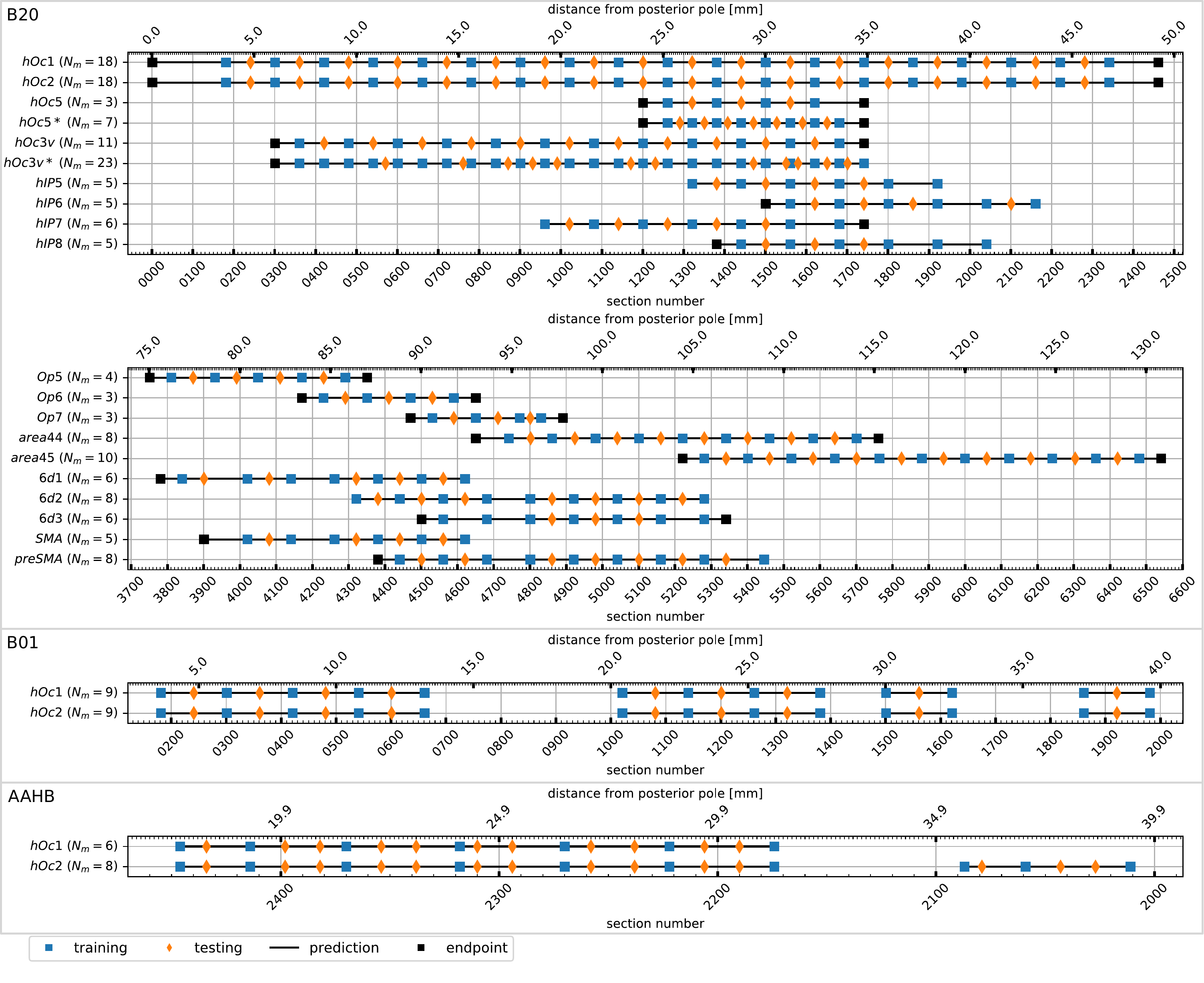}
	\caption{Training and test sections from available annotations across stacks of histological sections.
		Consecutive pairs of training sections (blue squares) induce one local segmentation model.
		For example, model \model{181}{301}{\bb-\vone} was trained on sections 181 and 301 of dataset \bb, segments area \vone in the full interval \interv{182}{300}, and was tested on section 241.
		$N_m$ denotes the number of trained local segmentation models.
		$\ast$~marks experiments performed with a smaller training interval.
	}
	\label{fig:training_section_overview_all}
\end{figure*}

We trained local segmentation models for 18 cytoarchitectonic areas in \bb and two areas in each of \bone and \ba.
\fref{fig:training_section_overview_all} gives an overview of sections used for the individual areas.
Most local segmentation models were trained on two training sections with annotations at $\sim 2.4\mm$ distance, corresponding to $\sim 120$ sections for \bb and \bone and $48$ sections for \ba.
Additional local segmentation models with a reduced interval size of $60$ sections ($1.2\mm$) were trained for areas \vthree and \vfive to account for highly variable morphology (\vthree, see \fref{fig:challenges}, C-F) and small area size (\vfive).
For \bone and \ba, local segmentation models were trained only for ranges of sections where annotations were available at the required interval.
Segmentations of the outer most parts of cytoarchitectonic areas which were not enclosed by training sections (i.e. sections 1 to 181 for \vone in \bb) were processed using the closest available local segmentation model.
For example, model \model{181}{301}{\bb-\vone} was also applied to the section interval \interv{1}{181}.

\subsection{Neural network architecture}%
\label{sub:network_architecture_for_local_segmentation_models}

\begin{figure*}[t]
	\centering
	\includegraphics[width=1.0\linewidth]{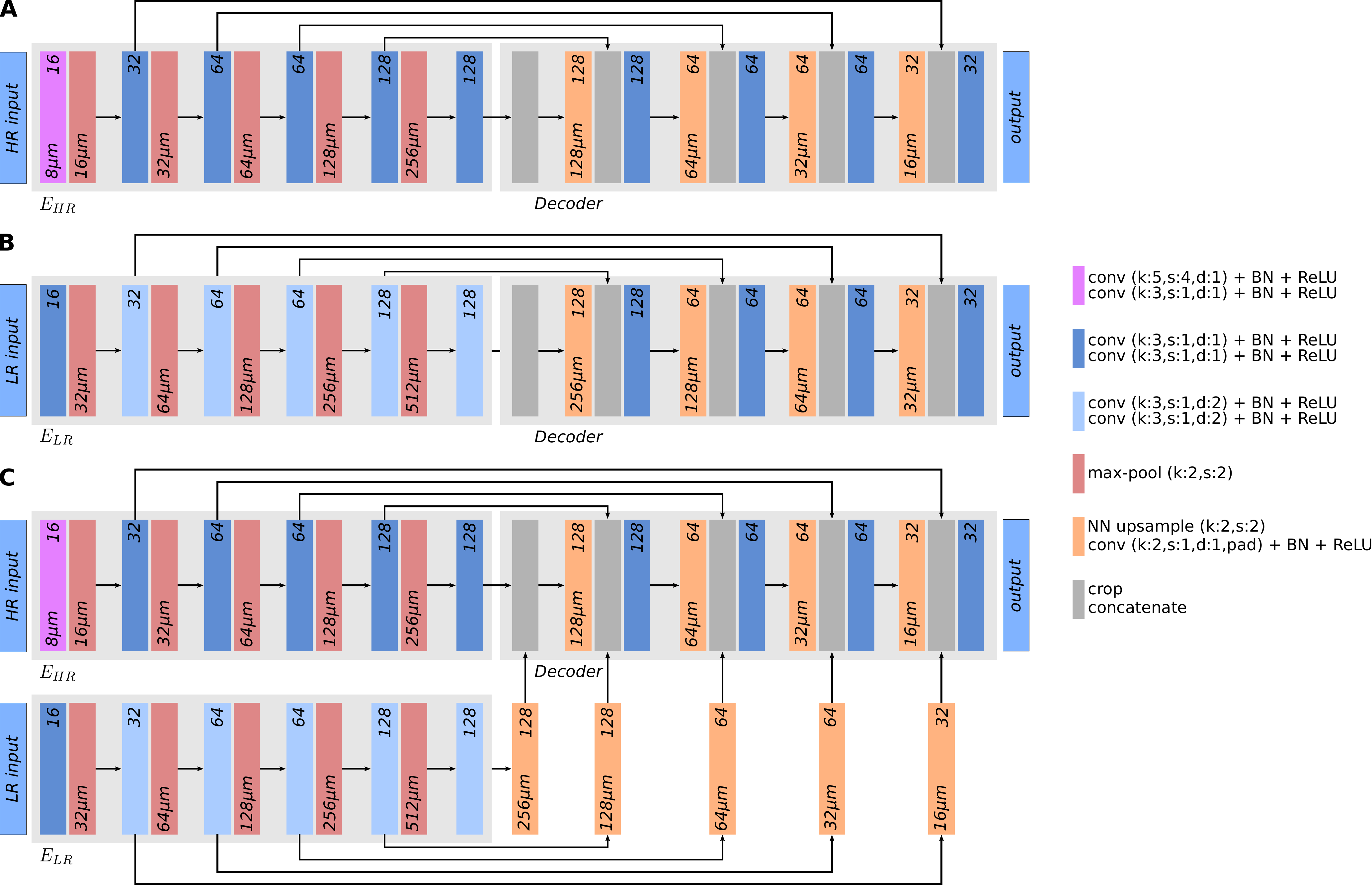}
	\caption{Illustration of investigated neural network architectures.
		\emph{A)} High resolution architecture (\hr) from \cite{Spitzer2017}{}, which can capture fine-grained microstructural textures.
		\emph{B)} Low resolution architecture (\lr), which can capture macroscopic tissue features.
		\emph{C)} Proposed multi-scale architecture (\ms) to capture both fine and coarse grained tissue features.
		\hre is pre-initialized with weights of the self-supervised network proposed in \cite{Spitzer2018}{}.
		Numbers at the top of each block denote the number of filters used in the convolutional layers of this block.
		Numbers at the bottom denote the physical output spacing in $\micron$ per pixel for layers which change the physical spacing of the features.
	}
	\label{fig:network_architecture}
\end{figure*}

For local segmentation models, the modified U-Net architecture~\cite{Ronneberger2015}{} proposed by~\cite{Spitzer2017}{} was extended into a multi-scale neural network model (\fref{fig:network_architecture}, C).
U-Nets have proven to be very powerful for many applications in biomedical image segmentation (e.g.~\citep{Cicek2016, Milletari2016}{}).
They consist of an encoder and decoder branch, which are linked by skip-connections between layers of corresponding spatial resolution to allow recovery of fine-grained details during upsampling.
To show the benefit of using a multi-scale variant of U-Nets, three network variants were used: A high-resolution encoder network (\hr), a low-resolution network (\lr), and a combined multi-scale architecture (\ms).

\paragraph{High Resolution Encoder architecture (\hr)}%
\label{par:high_resolution_encoder_architecture_hr}

The architecture proposed in \cite{Spitzer2017}{} was used as base architecture (\fref{fig:network_architecture}, A).
A \textit{high resolution encoder} \hre receives high resolution input patches with a size of $2025 \times 2025$ pixels at $2\micron$ pixel resolution ($4.05\times 4.05\mm^2$) and enables recognition of fine-grained microstructural textures.
It consists of six convolutional blocks, with the number of filters set to $\set{16, 32, 64, 64, 128, 128}$ respectively.
All but the last block are followed by a max-pooling operation with pool size 2 and stride 2.
The first layer of the first block in \hre uses a filter size of 5 and a stride of $4$, which increases the receptive field while keeping memory consumption and computational effort tangible.
All remaining convolutional layers of \hre use a kernel size of 3 and stride 1.
Weights of \hre were pre-initialized from the trained self-supervised network proposed in \cite{Spitzer2018}{}.
The decoder consists of four convolutional blocks with the number of filters set to $\set{128, 64, 64, 32}$ respectively.
Each block is preceded by an upsampling block, which consists of a nearest neighbor upsampling with kernel size 2 and stride 2, followed by a zero-padded convolutional layer with kernel size 2 and stride 1.
All convolutional operations in the network are followed by batch normalization~\cite[]{Ioffe2015} and \ac{ReLU} non-linearity.

\paragraph{Multi-scale network architecture (\ms)}%
\label{par:multiscale_network_architecture_ms}

The multi-scale network architecture was obtained by attaching a \textit{low resolution encoder} \lre as a second branch to \hr, which receives lower resolution image patches with a size of $682 \times 682$ pixels at $16\micron$ pixel resolution ($10.912\times 10.912\mm^2$), centered at the same location as \hre patches.
This branch allows to learn features at the scale of local cortical folding patterns.
Although such macroscopic features are not generally representative of cytoarchitecture in human brains, as they vary largely between individuals~\cite{Amunts2015}{}, they are appropriate in the present setting due to the locality of the network models.
\lre is based on \hre, and composed of six convolutional blocks with the same number of filters as \hre.
All convolutional filters use a filter size of 3 and a stride of 1.
Convolutional layers in the first block use a dilation rate of 1, while all other convolutional layers within \lre use a dilation rate of 2 to enlarge the receptive field.

\paragraph{Low Resolution Encoder architecture (\lr)}%
\label{par:pure_low_resolution_encoder_lr}

The third architecture is based on \hr, but replaces the encoder \hre with \lre (\fref{fig:network_architecture}, B).
By design, this model can only recognize macroscopic tissue features, and no detailed cytoarchitectonic properties at the level of cell bodies.

\subsection{Training strategy}%
\label{sub:training_strategy}

Stochastic gradient descent with Nesterov momentum~\citep{Sutskever2013} was used as optimizer for training the neural network models.
Training was performed for $3000$ iterations.
The learning rate was initially set to $0.01$ and decreased by a factor of $0.5$ after $1000$, $1400$, $1800$, $2200$ and $2600$ iterations.
Momentum was set to $0.9$.
Categorical cross-entropy with a weight decay of $0.0001$ was used as loss function.

\paragraph{Background class labels}%
\label{par:background_class_labels}

\cite[]{Spitzer2017} reported convergence problems when training models with a single background class that includes both white and gray matter components, resulting in a mix of tissue parts with very high and very low similarity to the target area under the same classification label.
Thus, the general background class was split into separate labels for gray matter (\cor) and white matter (\wm), resulting in a semantic segmentation problem with the four classes \bg, \wm, \cor, and the target area \apl.
For splitting the background class into \wm and \cor, different strategies were used for each dataset:
\begin{enumerate}
	\item For \bb, a volumetric tissue classification presented in~\cite{Lewis2014}{} was projected onto the 2D histological sections using transformations provided by the authors of~\cite[]{Amunts2013}.
	\item For \bone, the gray white matter segmentation described in \cite[]{Spitzer2017} was used.
	\item For \ba, the respective delineations available from the Allen ontology~\cite[]{Ding2016} were used.
\end{enumerate}

\paragraph{Patchwise training}%
\label{par:patchwise_training}

The full resolution scans of the whole-brain sections are by far too large to be used for training.
Thus, a patchwise training procedure as also proposed in~\cite{Ronneberger2015, Spitzer2017, Spitzer2018}{} was employed.
However, due to the locality of local segmentation models, patches were sampled only in the direct proximity of the target brain area \apl, to effectively teach the models to distinguish \apl from its immediate surroundings.
Only pixels with a distance of $5\mm$ or less to any pixel annotated as \apl were considered as potential center points for training patches.

\paragraph{Data augmentation}%
\label{par:data_augmentation}

The following data augmentations were employed to simulate most frequently observed variations in the data:
Both at test and training time, images were rotated by multiples of 90 degrees so that the $y$ axis of coronal sections matches approximately the cranial
direction.
Random rotation by an angle sampled from a uniform distribution with support $\interv{-45}{45}$ were applied to account for small differences in rotation angle.
Intensity variations were addressed by random pixel intensity augmentation with the function $f(x) = \alpha x^{\gamma} + \beta$.
The same intensity transformation is applied to all pixels of a training patch.
Parameters were chosen from uniform distributions with $\alpha \sim U\interv{0.9}{1.1}$, $\beta \sim U\interv{-0.2}{+0.2}$ and $\gamma \sim U\interv{0.8}{1.214}$.
The range of each parameter was empirically chosen to reflect natural variations occurring in the data.

\paragraph{Implementation}%
\label{par:implementation}

Training was performed on the supercomputer \textit{JURECA}\footnote{\url{https://www.fz-juelich.de/ias/jsc/EN/Expertise/Supercomputers/JURECA/JURECA_node.html}} at the Jülich Supercomputing Centre at Research Centre Jülich (JSC)~\cite{Krause2018}.
Each compute node was equipped with four NVidia K80 GPUs with $12$ Gigabyte of VRAM, 2 Intel Xeon E5-2680 v3 Haswell CPUs ($12$ $2.5$ GHz cores with hyperthreading each) and $128$ Gigabyte of RAM~\cite{Krause2018}{}.
Training of one model occupied one GPU node, using all 4 GPUs and all 48 threads.
Of the available 48 threads, 4 were assigned to one GPU each to coordinate the training process, while the remaining 44 threads read training patches from disk in a streaming fashion, applied data augmentation and sent data to the training threads.
Inter-process communication was implemented based on \ac{MPI} using \textit{mpi4py}~\cite{Dalcin2011}.
Training was implemented using \textit{TensorFlow}~\cite{Abadi2016}{}.
Distributed training was performed using \textit{Horovod}~\cite{Sergeev2018} and synchronous distributed stochastic gradient descent.
Batch size was set to $16$ image patches per GPU, resulting in a total effective batch size of $64$ patches per iteration.
The linear learning rate scaling rule for distributed training proposed in~\cite[]{Goyal2017} was employed, scaling the learning rate by the number of GPUs\footnote{Since we use a relatively small number of employed GPUs however, we do not apply the initial learning rate warm up phase described in~\cite{Goyal2017}{}.}.
Batch normalization statistics were computed independently for each GPU and not averaged during training.
Software code is publicly available~\footnote{Code available at \url{https://jugit.fz-juelich.de/c.schiffer/atlas}}.

\subsection{Web-based interactive workflow for efficient \mbox{cytoarchitectonic} mapping}%
\label{sub:web_based_interactive_workflow_for_efficient_cytoarchitectonic_mapping}

The proposed workflow was implemented as an interactive web application to provide direct user control over the segmentation workflow through a web browser\footnote{Code available at \url{https://jugit.fz-juelich.de/c.schiffer/atlasui}}.
The application allows entering annotations in a sparse set of reference sections, controlling the training workflow on a remote cluster, and efficiently inspecting predicted segmentations in the complete stack of histological sections.
It does not require in-depth expertise in Deep Learning and/or batch computations.
Technically, it is designed as an extension of the web based annotation tool \emph{microdraw}~\footnote{\url{http://microdraw.pasteur.fr}}, combined with a novel backend service that controls data exchange and job supervision on an ssh-accessible compute cluster (here the JURECA system at JSC in Jülich).
The workflow typically iterates through the following steps:
\begin{enumerate}
	\item The user enters annotations for a target brain area $\apl$ in two tissue sections $s_1$ and $s_2$, enclosing a local stack interval of $\approx100$ sections using \textit{microdraw}, and this way defines a local segmentation model \model{s_1}{s_2}{\apl} (\sref{sub:local_segmentation_models}).
	\item A training task for the local segmentation model is submitted as a job to a GPU cluster at the push of a button, using default parameters (\sref{sub:training_strategy}).
		It does not require any further configuration.
		Training typically takes 70 minutes on one compute node of the JURECA supercomputer.
		Multiple jobs can be submitted in parallel, if the cluster allows.
	\item After training, predictions for all sections in the interval are automatically generated.
		For a large area like \vone{}, this takes approximately 30 minutes for 120 sections.
		Computed segmentations are automatically displayed in the web frontend once they become available.
		Data synchronization between the web server and compute nodes is handled by the backend service.
	\item After inspecting the segmentation quality, the user can choose to enter additional training data, either reducing the size of the current interval or initiating the next interval in the stack.
\end{enumerate}

\subsection{Validation framework and strategy}%
\label{sub:validation_framework}

Additional sections with annotations in between the training sections were used for validating performance of local segmentation models on sections that were not seen during training (orange diamonds in \fref{fig:intervals}).
Segmentations of these test sections were quantitatively evaluated using the F1 score (also known as Dice score or S\o rensen-Dice index), computed as the harmonic mean of precision of recall.
Auxiliary labels added to ensure convergence (\sref{par:background_class_labels}) were excluded from F1 score calculation, as the focus lies on segmentation performance for target area \apl.

Similar to the proximity sampling strategy employed for training (\sref{par:patchwise_training}), segmentations on sections not seen during training were only created and evaluated in the approximate region containing \apl on the respective sections.
These approximate regions were determined by projection of the closest reference annotations for \apl to the image in question using conventional linear image registration based on robust image features as in~\cite[]{Dickscheid2019}.

The benefit of a multi-scale architecture was investigated by training separate local segmentation models with neural network architectures \hr, \lr and \ms for all areas in \bb.
For \hr and \ms, the high resolution encoder \hre was initialized with the weights of the network from~\cite[]{Spitzer2018}.
Furthermore, the performance of multiple local segmentation models, each trained on a local subset of sections as described in \sref{sub:local_segmentation_models}, was compared to the performance of one single model trained on all annotations available for a target area \apl in the following way:
For each target area in the \bb dataset, one model was trained using the union of all training sections of the local segmentation models (blue squares in \fref{fig:intervals}), using the same training strategy as for local segmentation models.
As the models are not local and cannot rely well on tissue morphology, this comparison was only performed using the \hr architecture, again pre-initializing the high-resolution encoder \hre with weights from~\cite[]{Spitzer2018}.
Models trained on the full stack instead of local intervals are denoted as \hrall.

The robustness of the proposed method against intersubject variability in brain structure and differences in staining protocols was investigated by training local segmentations models (with \ms architecture) for areas \vone and \vtwo in datasets \bone and \ba.

\subsection{Generating high-resolution 3D cytoarchitectonic maps in the BigBrain dataset}
\label{section:bigbrain_3d}

Non-linear transformations described in~\cite[]{Amunts2013, Omidyeganeh2020} from 2D histological sections into 3D reconstructed space available for the BigBrain dataset~\cite[]{Amunts2013} were used to generate 3D maps for areas \vone, \vtwo, \vthree and \vfive from 2D segmentations produced by our method.
Segmentations were obtained using the workflow described in \sref{sub:local_segmentation_models} and checked for quality by an expert (e.g. plausibility and consistency across consecutive sections).
For areas \vthree and \vfive, results of segmentation models trained with a training interval size of $1.2\mm$ were used for reconstruction (marked with $\ast$ in \fref{fig:training_section_overview_all}).
Between $8\%$ (\vthree) and $23\%$ (\vone) of sections containing the investigated areas were not used for reconstruction due to histological artifacts (e.g. resulting from long-term storage or staining inhomogeneities).
Segmentations that passed the quality check were transformed into the 3D reconstructed space.
Excluded sections were replaced by interpolations from neighboring sections, using Laplacian fields as proposed in~\cite[]{Schober2016}.

\begin{figure}
	\centering
	\includegraphics[width=1.0\linewidth]{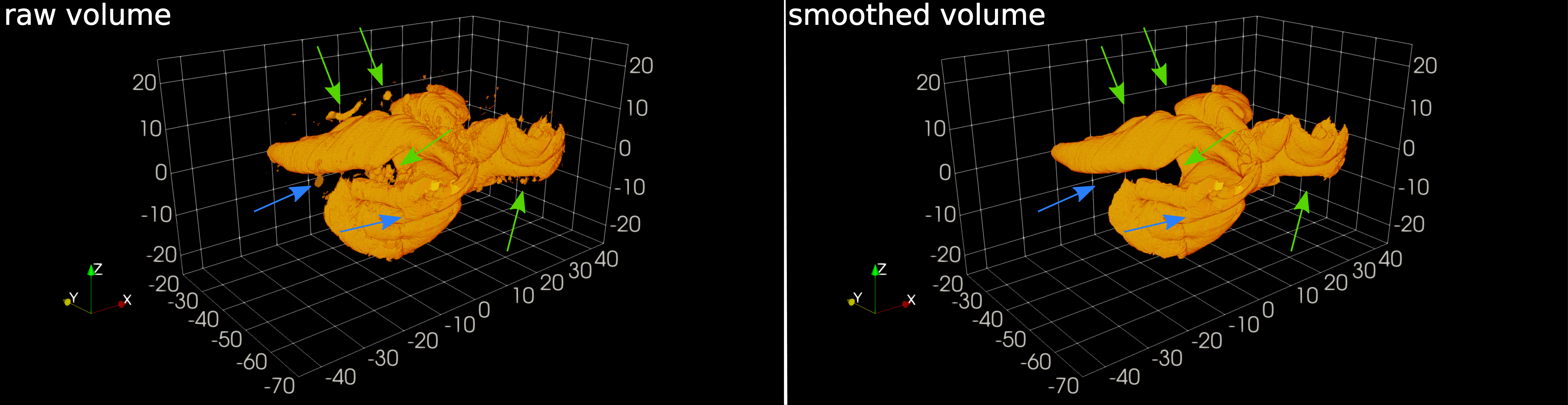}
	\caption{Effects of median filtering and connected component filtering using the example of \vone.
		Median filtering smooths the volume and removed small errors originating from registration errors or incorrect predictions (blue arrows).
		Filtering of small connected components removed small clusters of false positives from the volume (green arrows).
		Axes x, y and z correspond to left-to-right, posterior-to-anterior and ventral-to-dorsal directions, respectively.
		Axis labels are specified in $\mm$ and correspond to positions in the 3D reconstructed BigBrain space.
	See \fref{fig:reconstruction_images} for more images of \vone from different viewing angles.}
	\label{fig:connected_components}
\end{figure}

Resulting 3D maps were smoothed using a median filter with kernel size $11\times 11\times 11$ pixel to compensate for small artefacts.
The size of the filter was chosen to match the expected precision of annotations at boundaries (not higher than $100\micron$), translating to 5 voxels at the target resolution of $20\micron$.
Furthermore, connected component analysis on the smoothed volume was performed to determine and remove spurious false positive predictions outside the target area, relying on the assumption that cytoarchitectonic areas are continuous in 3D.
Only components with a minimum volume of $27\mm^{3}$ ($3\mm \times 3\mm \times 3\mm$) were kept.
Effects of median filtering and connected component filtering are illustrated in \fref{fig:connected_components}.

To assess the improvement in 3D consistency and anatomical plausibility gained by the proposed workflow, a reference reconstruction of area \vone was computed, which performs a direct 3D interpolation between reference annotations obtained by GLI mapping.
This reference reconstruction does not use the local segmentation models, and relies only
on reference annotations and 3D reconstruction.
It was computed by transforming the annotations of the training sections (blue squares in \fref{fig:training_section_overview_all}) into the 3D reconstructed space, and filling the gaps by Laplacian field interpolation~\cite{Schober2016}{}.

The anatomical consistency of 3D reconstructed maps was further evaluated by computing their volume and surface area, which were then compared to reference values from~\cite[]{Amunts2000} .
The volume of each area was computed by counting the total number of labeled voxels and multiplying the result by the physical size of each voxel.

The surface area was computed by first extracting a closed surface mesh of each area using the marching cubes algorithm~\cite[]{Lewiner2003}.
The subset of mesh vertices lying on the pial surface was then determined by including all triangles where the cortical depth~\cite[]{Bok1929} was smaller than $0.25$.
To obtain the cortical depth of each mesh vertex, the procedure described in~\cite{Leprince2015}{} was applied to the cortical ribbon defined by the gray and white matter segmentation provided with the BigBrain model~\cite{Lewis2014}{}.
The result was a volumetric dataset with voxels in the white matter labelled 1, voxels outside the brain labelled 0, and voxels inside the isocortex labelled with values between 0 and 1, representing their cortical depth according to the equivolumetric model~\cite{Bok1929}{}.
Cortical depths of mesh vertices were then looked up in this volume.
Finally, the surface area of the pial surface for each cytoarchitectonic area was computed by summing up the area of all triangles associated to the pial surface.

Both volume and surface area measurements were corrected for tissue shrinkage~\cite[]{Amunts2000}.
The volume-based shrinkage factor for \bb has been determined in~\cite{Amunts2005}{} based on the fresh weight and the volume after histological processing as $f_V = 1.931$.
From this, an area-based (2D) shrinkage factor of $f_A=f_V^{2/3} = 1.551$ was derived.

\section{Results}%
\label{sec:results}

\begin{figure}
	\centering
	\includegraphics[width=1.0\linewidth]{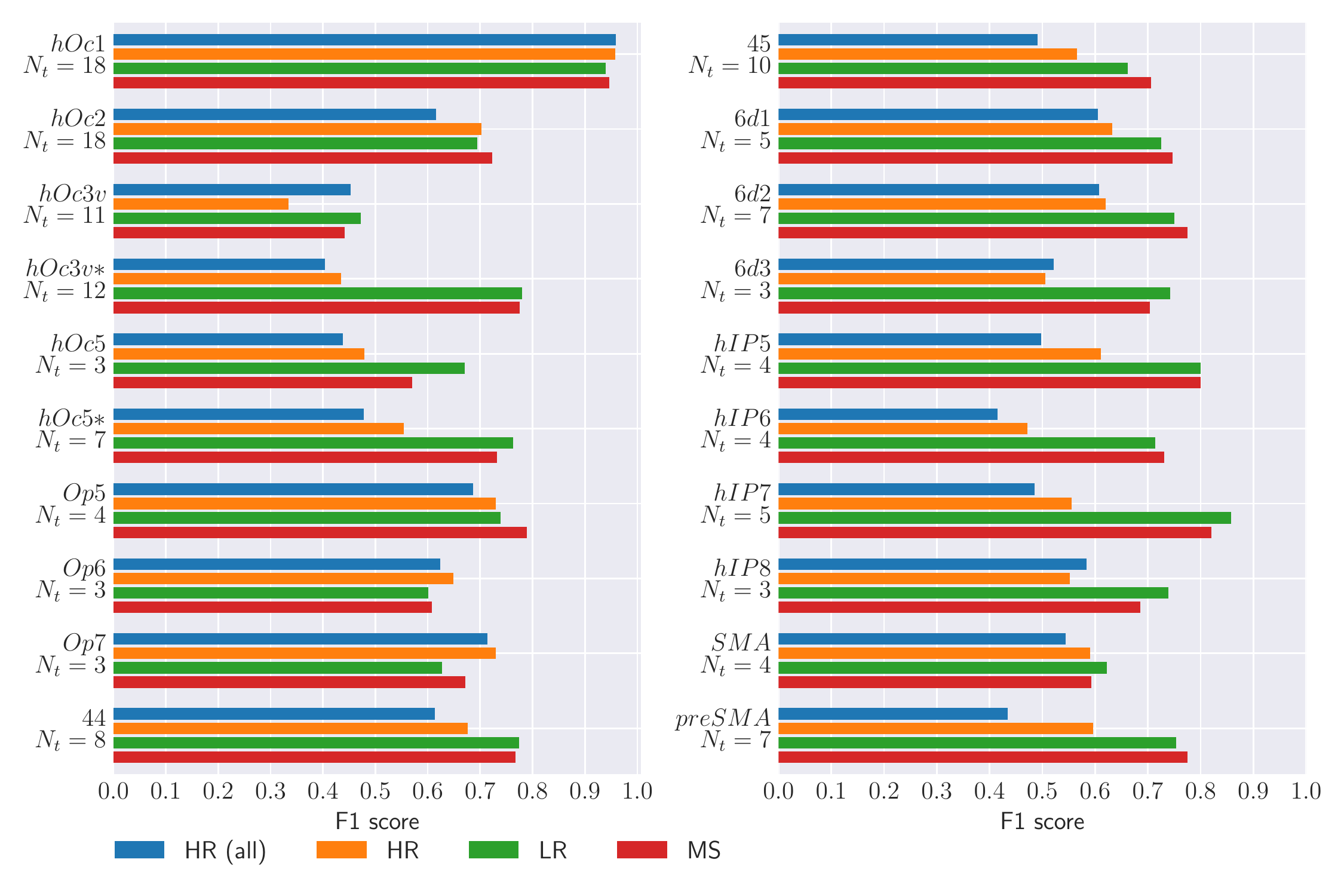}
	\caption{Median F1 scores for \hr, \hrall, \lr and \ms per investigated brain area in dataset \bb.
		$N_t$ denotes the number of test sections for which F1 scores were computed for a particular area.
		$\ast$~indicates where training of local segmentation models was performed with reduced distance between training sections.
		Higher values denote better performance.
	}
	\label{fig:bar_f1_score_model_comparison}
\end{figure}

The performance of the models \hr, \hrall, \lr and \ms differed between each other.
Lowest mean and median performance were obtained by \hrall, followed by \hr.
Both \lr and \ms resulted in considerably higher mean and median F1 scores than the high resolution architectures, with lower standard deviations.
Highest mean and median performance was obtained by \ms (\tref{table:model_comparison_statistics}, \fref{fig:bar_f1_score_model_comparison}).

\begin{SCtable}
	\centering
	\caption{F1 score statistics computed across all areas and test sections in the \bb dataset obtained by the different network architectures \hrall, \hr, \lr and \ms.
		Higher mean/median values and lower standard deviation mean better performance.
	}
	\begin{tabular}{lrrr}
\hline
 model     &   median &   mean &    std \\
\hline
 \hr (all) &   0.5319 & 0.5680 & 0.2075 \\
 \hr       &   0.6294 & 0.6130 & 0.2105 \\
 \lr       &   0.7439 & 0.7036 & 0.1865 \\
 \ms       &   0.7469 & 0.7200 & 0.1825 \\
\hline
\end{tabular}
	\label{table:model_comparison_statistics}
\end{SCtable}

Performance also differed significantly between brain areas.
All architectures show comparably good performance for \vone.
For most areas however, \lr and \ms achieved considerably higher performance than \hr and \hrall.
For areas \vthree and \vfive, where additional models were trained with reduced distance between training sections (indicated by $\ast$ in \fref{fig:bar_f1_score_model_comparison}), performance of \lr and \ms increased considerably when decreasing the distance between training sections, while only minor improvements were observed for \hr and \hrall.

\begin{figure*}
	\centering
	\includegraphics[width=1.0\linewidth]{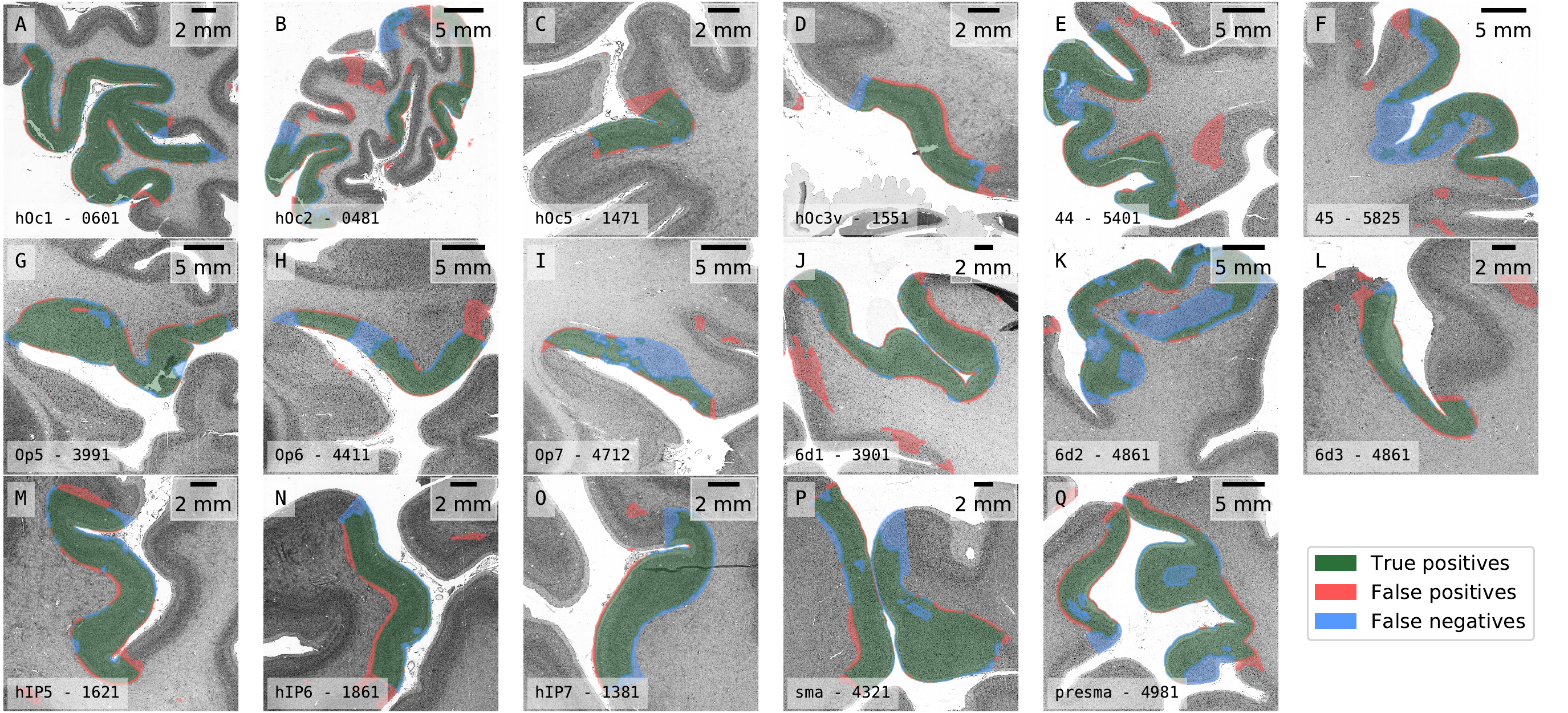}
	\caption{Example image patches and corresponding model predictions extracted from test sections of \bb segmented using the proposed \ms architecture.
		One image patch is shown for each investigated cytoarchitectonic area.
		Colors green, red and blue indicate true positive, false positive and false negative predictions, respectively.
	}
	\label{fig:example_patches}
\end{figure*}

\begin{SCfigure}
	\includegraphics[width=0.7\linewidth]{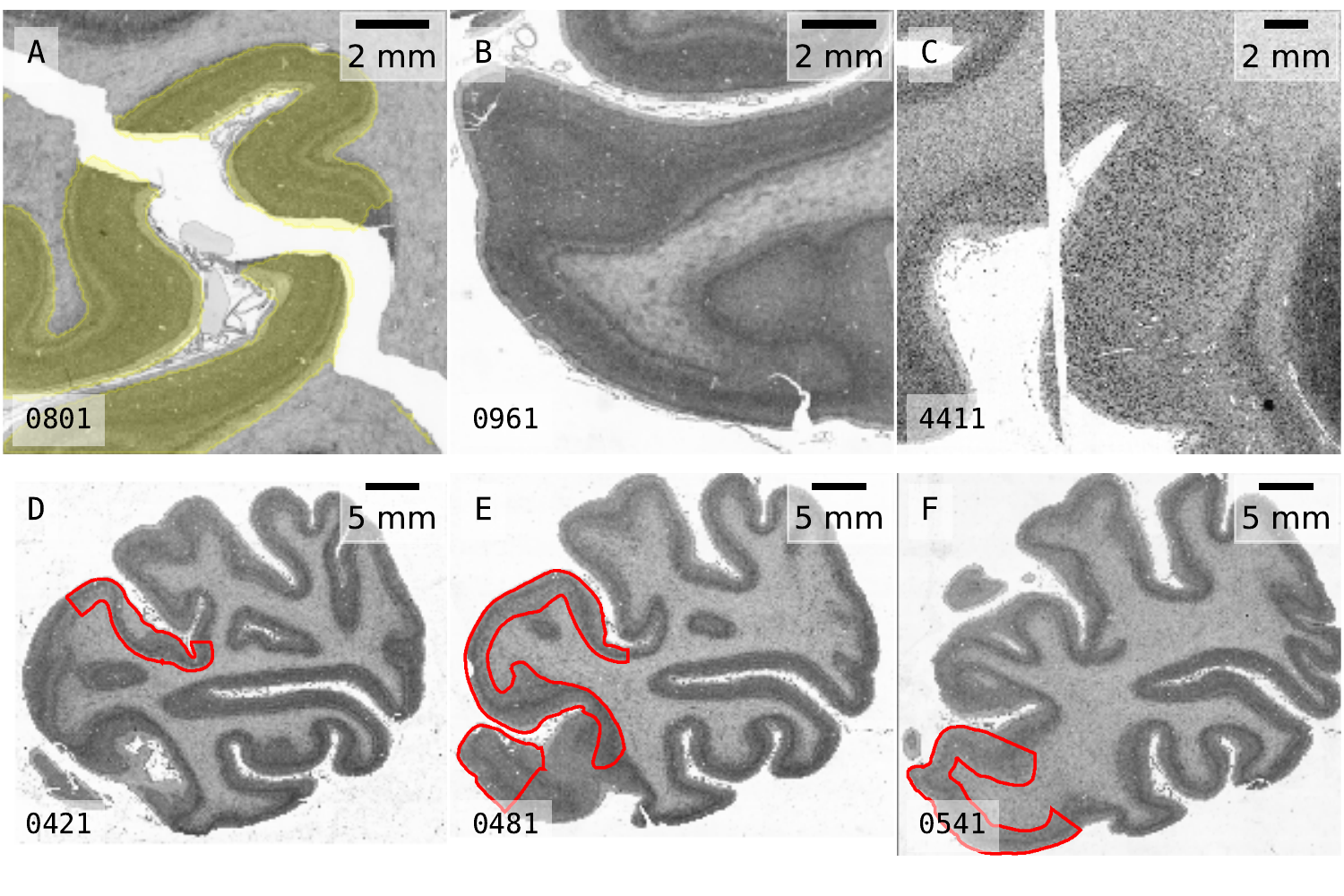}
	\caption{Image patches extracted from \bb showing common challenges encountered during manual and automated cytoarchitectonic mapping.
		\emph{A:}~Mechanical damages resulting from histological processing. Prediction for \vone shown in yellow demonstrate handling of larger mechanical damages.
		\emph{B:}~Region where the cutting angle is highly oblique, leading to partial or full occlusion of cortical layer structure (\textit{oblique cuts}).
		\emph{C:}~Mechanically damaged and obliquely cut tissue.
		\emph{C-F:}~Example illustrating highly variable morphology of area \vthree (highlighted in red) across 120 histological sections in \bb.
	}%
	\label{fig:challenges}
\end{SCfigure}

Representative image patches segmented by the \ms architecture for each investigated area extracted from test sections of \bb are shown in \fref{fig:example_patches}.
True positive, false positive and false negative predictions are indicated in green, red and blue, respectively.
A large share of incorrectly classified pixels belonged to cortical regions with highly oblique cutting angles (\fref{fig:challenges} B, C).
While large rifts tended to be excluded from the prediction (\fref{fig:challenges}, A), smaller rifts or tissue foldings were correctly segmented as surrounding area (\fref{fig:example_patches}, A, D, E, G, O).

\begin{figure}[t]
	\centering
	\includegraphics[width=1.0\linewidth]{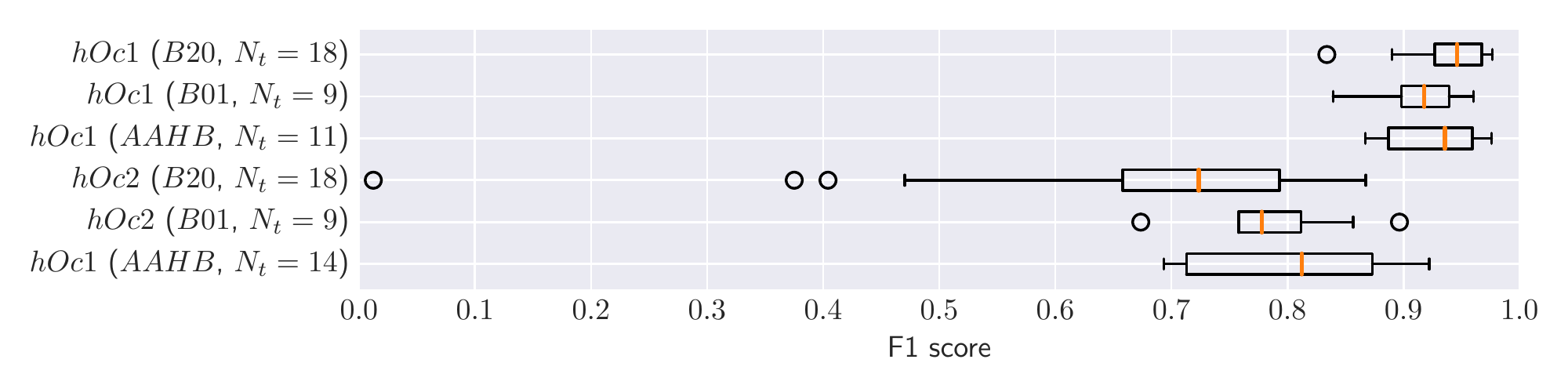}
	\caption{F1 scores for segmentations of \vone and \vtwo obtained by the \ms architecture on test sections of datasets \bb, \bone and \ba.
		$N_t$ denotes the number of test sections for which F1 scores were computed for a particular area.
	}
	\label{fig:boxplot_f1_score}
\end{figure}

\begin{figure*}
	\centering
	\includegraphics[width=1.0\linewidth]{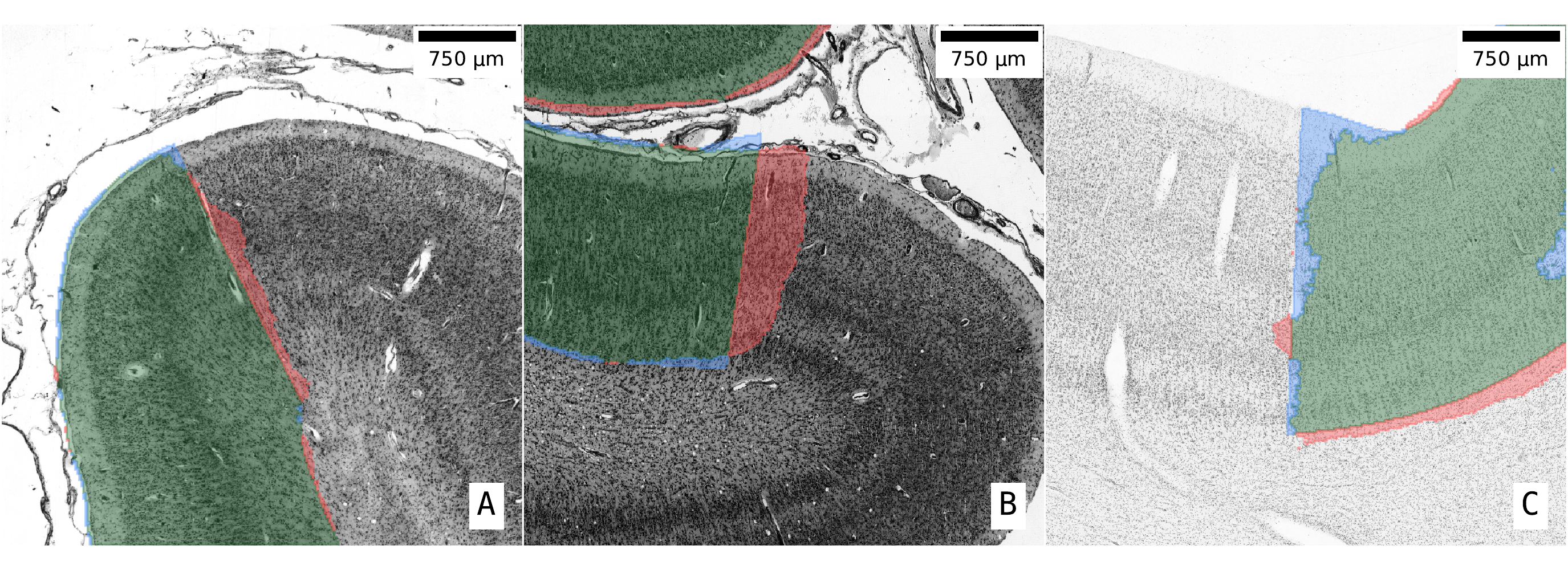}
	\caption{Example patches and typical segmentation results extracted from test sections in datasets \bb (A), \bone (B) and \ba (C).
		All three patches show the segmentation of \vtwo obtained by a local segmentation model with \ms architecture.
		Patches were extracted at the border between \vone and \vtwo and in comparable regions of the respective brain.
		Colors green, red and blue indicate true positive, false positive and false negative predictions, respectively (see also legend in \fref{fig:example_patches}).
	}
	\label{fig:example_patches_brain_comparison}
\end{figure*}

Scores obtained for areas \vone and \vtwo were overall consistent across different brain samples~(\fref{fig:boxplot_f1_score}).
In all three cases, scores obtained for \vtwo were lower compared to \vone.
Lowest median F1 score for \vtwo was obtained for \bb, along with an increased variance.
Example patches showing the border between \vone and \vtwo on test sections extracted from approximately identical brain regions in the three datasets are shown in \fref{fig:example_patches_brain_comparison}.

\begin{figure}
	\centering
	\includegraphics[width=0.99\textwidth]{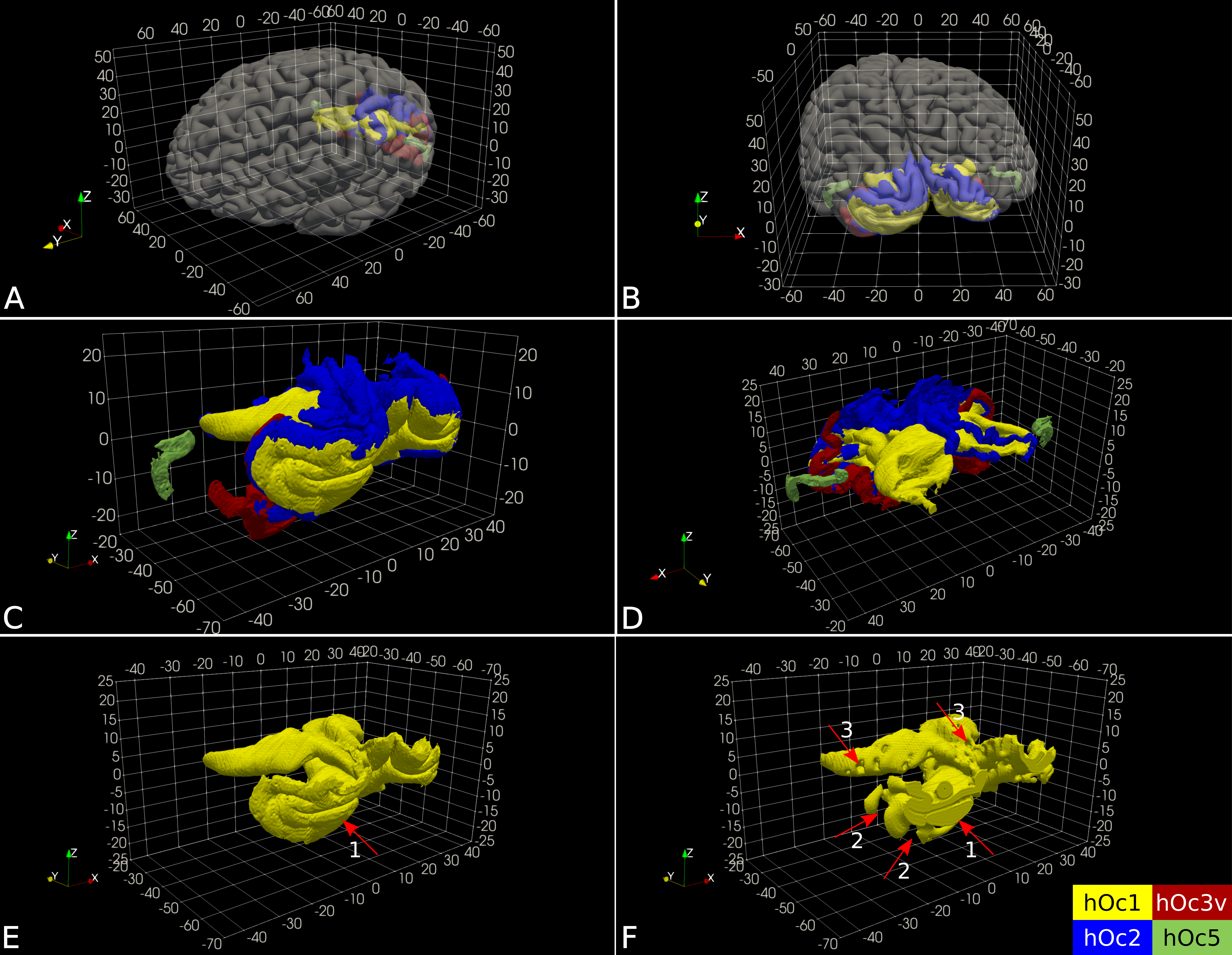}
	\caption{3D maps of visual cytoarchitectonic areas \vone (yellow), \vtwo (blue), \vthree (red) and \vfive (green), obtained by transforming the independent 2D segmentations generated by the proposed method into the 3D reconstructed space of the \bb dataset.
		\emph{A+B:} Spatial embedding of reconstructed areas into the 3D reconstructed BigBrain volume.
		\emph{C+D:} Detailed view of reconstructed cytoarchitectonic areas.
		\emph{E+F:} Comparison of \vone reconstructed based on our proposed method (E) and based on an interpolation between annotations in the reconstructed space, using Laplacian fields as proposed in \cite{Schober2016}{} (F).
		Arrows in F) mark example locations demonstrating shortcomings of the interpolation based reconstruction.
		\textbf{Axes x, y and z correspond to left-to-right, posterior-to-anterior and ventral-to-dorsal directions, respectively.
	Axis labels are specified in $\mm$ and correspond to positions in the 3D reconstructed BigBrain space.}}
	\label{fig:reconstruction_images}
\end{figure}

Locations, orientations and shapes of reconstructed 3D maps (computed using steps described in \sref{section:bigbrain_3d}) were anatomically plausible and consistent (\fref{fig:reconstruction_images}).
The 3D map of \vfive showed partially missing extremal ends along the posterior anterior axis.
Volume and surface estimates from the 3D maps reported in \tref{table:volumes_surface} corresponded well with the numbers reported in~\cite[]{Amunts2000}.
Surface areas of \vone, \vtwo and \vfive were largely confirmed with the reference values, as well as the volumes derived from automatic segmentations of areas \vone and \vtwo.
The reconstructed volume of area \vfive stood out by being considerably smaller than the reference volume.

\begin{SCtable}
	\centering
	\caption{Estimated volumes (in $\mm^3$) and surface areas (in $\mm^2$) of brain areas derived from the full 3D maps in the 3D reconstructed space of the \bb dataset.
		Reference mean $\mu$ and standard deviation $\sigma$ were computed based on male subjects from \cite{Amunts2000}.
		Shrinkage corrected of volumes and surface areas was performed using correction factors $f_V=1.931$ and $f_A=1.551$ respectively~\cite{Amunts2005}.
	}

\begin{tabular}{lrrlll}
\hline
 area      &   volume &   corrected & $\mu$   & $\sigma$   & z-score   \\
\hline
 \vone{}   &  9019.30 &    17416.27 & 18042.2 & 2464.39    & -0.25     \\
 \vtwo{}   &  6448.60 &    12452.26 & 12634.2 & 2862.84    & -0.06     \\
 \vthree{} &  1974.76 &     3813.26 & n.a.    & n.a.       & n.a.      \\
 \vfive{}  &   304.10 &      587.21 & 1144.4  & 406.53     & -1.37     \\
\hline\hline
 area      &   surface &   corrected & $\mu$   & $\sigma$   & z-score   \\
\hline
 \vone{}   &   6891.03 &    10685.76 & 12213.0 & 2225.55    & -0.69     \\
 \vtwo{}   &   6749.64 &    10466.52 & 10390.4 & 2925.37    & 0.03      \\
 \vthree{} &   2142.04 &     3321.62 & n.a.    & n.a.       & n.a.      \\
 \vfive{}  &    319.79 &      495.89 & 450.2   & 135.92     & 0.34      \\
\hline
\end{tabular}
	\label{table:volumes_surface}
\end{SCtable}


Comparison of corresponding 3D reconstructions of area \vone (\fref{fig:reconstruction_images} E vs.~ \fref{fig:reconstruction_images} F) showed that the proposed approach provided anatomically more consistent results than direct spatial interpolation of GLI-based annotations, while both build on the same annotation effort.
3D interpolation produced abrupt transitions in anterior-posterior direction (\fref{fig:reconstruction_images}, F, 1) and only captured structures already contained in the reference annotations, leading to inconsistencies near fine-grained morphological structures (e.g.~\fref{fig:reconstruction_images} F, 2 and 3).
The proposed method often produced reasonable segmentations for sections outside the training interval (\fref{fig:reconstruction_images}, E, 1), which interpolation cannot provide by definition.

\section{Discussion}%
\label{sec:discussion}


In this work, we proposed a novel Deep Learning based workflow to create segmentations of cytoarchitectonic areas in large series of histological human brain sections using only a limited set of manually created annotations.
We evaluated this approach across different cytoarchitectonic areas, brain samples and staining protocols.
As a concrete use case, we then applied it to create high-resolution 3D maps of areas \vone, \vtwo, \vthree and \vfive in the BigBrain~\cite[]{Amunts2013}.

\subsection{Quality of derived 3D maps in the BigBrain}%
\label{par:high_resolution_3d_maps_in_the_bigbrain}

The proposed method produced 3D maps with a high degree of anatomical consistency and identified cytoarchitectonic areas precisely in the histological brain sections.
Partially missing extremal ends remain a challenge, as seen in anterior-posterior direction of \vfive.
Such parts are often difficult to identify even using manual methods.
Therefore, training data for such extremal ends is difficult to obtain.
The segmentation of extremal ends could potentially be addressed by providing additional GLI-based mappings (at the cost of additional annotation effort), or by an explicit shape-based inference step on top of the pixel segmentation.
The 3D map of \vone created with the proposed method is superior to the map obtained by direct spatial interpolation between GLI-based annotations.
Methods based on 3D interpolation inherit any error in the alignment of consecutive sections, making them inappropriate for stacks with only linear or no 3D reconstruction.
The proposed method does not assume any prior 3D reconstruction - in fact its outputs might be used to guide image registration with landmarks.

\subsection{Practical usefulness of the implemented workflow}

The presented method showed good robustness against intersubject variability and different histological processing protocols.
Thus it largely overcomes the need for brain or area specific parameter adjustments, which makes it well suited to be used as a self-contained tool for neuroscientists.
Consequently, it was possible to implement it into a web application that provides a practical mapping workflow for end users from different disciplines.
The web application is currently used by five neuroscientists in our institute for their research projects, without requiring support from a computer scientist.
The interactive workflow enables efficient mapping of brain areas across full series of histological sections, en par with high throughput microscopy. Such efficiency of mapping was previously impossible in our experience.
To give a concrete example, we consider that a trained expert typically needs 30-60 minutes to identify cytoarchitectonic borders for one cortical area on a single tissue section.
Using the established GLI-based mapping approach (see \sref{sub:observer_independent_mapping_according_to_gli_profiles}), this would translate to an approximate effort of 150 work days (8 hours per day) to map \vone across the whole stack of 2461 sections.
In comparison, the proposed method required annotation of only 18 sections to generate precise segmentations of the complete stack, corresponding to approximately 9 working hours.
Altogether, including quality checks and computations, the presented workflow allows precise mapping of a large brain area in the order of 1-2 weeks - a task that would require almost a year of work with previously established methods.

\subsection{Ability to distinguish higher associative areas}

In contrast to primary areas such as the primary visual cortex \vone{}, so called higher associative areas have a less distinct cytoarchitecture, and less prominently differ from their neighbouring areas.
Such observation lead Bailey and von Bonin to the conclusion that it is almost impossible to reliably distinguish such areas from each other, and to define borders between them~\cite{Bailey1951}.
This view is not supported any more due to the possibility to identify cytoarchitectonic borders in a reliable and reproducible way (for an overview see~\cite{Zilles2010}).
However, the fact that intersubject differences between identical areas of different brains may exceed cytoarchitectonic differences between two neighboring areas in one and the same brain creates challenges for modern brain mapping~\cite{Amunts1999}.

\cite[]{Atzeni2018} also addressed automated mapping of histology.
They segmented brain structures in a serial stack of human brain sections from the Allen Human Brain Atlas~\cite{Ding2016}{} (dataset \ba used in our experiments).
They used annotations from \cite[]{Ding2016} on a small set of sections at regular intervals, in order to train a probabilistic model that combines multi-atlas segmentation with a \ac{CNN} to segment the remaining sections.
Compared to the present work however, their approach is restricted to brain structures that can be recognized at a resolution of $250\micron$.
The authors confirm in their paper that more subtle classes, in particular subdivisions of the isocortex, introduce excessive noise with their approach.
The method presented here segmented both \vone and \vtwo in the same dataset with high accuracy by including more fine-grained texture features into the models, thus going clearly beyond this restriction.

\subsection{Benefits of using local segmentation models}%
\label{par:methodological_discussion}

Previous work on automatic cytoarchitectonic brain mapping based on machine learning emphasized that it is essential to find efficient strategies for exploiting available training data and prior information, e.g.~by incorporating probabilistic priors from brain atlases~\cite[]{Spitzer2017} or self-supervised learning~\cite[]{Spitzer2018}.
The key idea of this work is to use multiple local segmentation models, each of which focuses on a spatially restricted subset of sections in a specific brain area.
It does not aim to learn a general classification model that captures multiple areas across many sections and brains.
This is an explicit design decision in order to maximize practical benefit.
The significantly improved performance of \hr compared to \hrall showed the benefit of using local segmentation models.
The local models can exploit specific cytoarchitectonic features in a particular interval of a specific brain area, while models that aim to generalize across areas and brains need to capture more general features.
Due to their strict locality, local models can further rely on morphological features like folding patterns which are not generally representative for cytoarchitecture, and vary largely between individuals~\cite[]{Amunts2015}.
Consequently, the incorporation of \lre in architectures \lr and \ms brought a significant performance gain compared to \hr and \hrall, which suggests that macroscopic tissue features are important for the local models.
Our intuition that a combination of microscopic and macroscopic tissue features should be optimal is confirmed by a Wilcoxon signed-rank test~\cite[]{Wilcoxon1945} ($p=0.0011$), that showed better performance of the multiscale architecture \ms compared to \lr{}.

A major advantage of the local segmentation models is the ability to flexibly adjust the distance between training sections to account for regions with particularly simple or complex properties.
This has been demonstrated for the challenging areas \vfive and \vthree, where a reduction of the distance between training sections from 120 ($2.4\mm$) to 60 ($1.2\mm$) improved precision to a satisfactory level while keeping the annotation effort tractable.
In a similar fashion, larger areas or areas with distinct cytoarchitectonic features (e.g. \vone) can be segmented with a coarser set of training sections, in this case reducing annotation effort.

On the downside of such local models, hyperparameter assessment (e.g. for learning rate or model architecture) is not straightforward when training multiple models on different training sets and evaluating them on individual test sets.
Model performance needs to be evaluated across several areas, sections and brains, which can be computationally expensive and lead to a slow development process.

\subsection{Mapping at highly oblique cutting angles}

Many of the remaining classification errors coincide with highly oblique cutting angles of the tissue.
As also reported in~\citep{Schleicher1999, Spitzer2017, Spitzer2018}{}, identification of cortical areas is almost impossible at such angles, because the laminar composition of the cortex is then almost invisible in the 2D section.
In such cases, experts would consult adjacent sections to identify areas, which the proposed method cannot do.
An extension of the method considering multiple adjacent sections for classification might be able to overcome this issue.

\section{Conclusion}%
\label{sec:conclusion}

A novel method based on Convolutional Neural Networks (\acp{CNN}) was introduced for automated mapping of cytoarchitectonic areas in large series of histological human brain sections.
Segmentation models were trained for segmentation of different cytoarchitectonic areas in histological stacks obtained from three different brain samples.
A key idea is to train separate local segmentation models based on annotations of one specific target area in only two training sections, to focus the learning process on microscopic and macroscopic tissue features close to the training sections.
After training, local segmentation models were able to accurately segment sections in between their respective training sections.
By concatenating results from multiple local segmentation models, segmentations for complete brain areas can be obtained.
The proposed method opens up new possibilities to map complete stacks of histological human brain sections in a highly automated fashion, and thus provides an important basis for building high resolution human brain maps for datasets like BigBrain.
To the best of our knowledge, the maps of areas \vone, \vtwo, \vthree and \vfive computed for the BigBrain model using this method are the first high-resolution 3D maps of human cytoarchitectonic areas created from full stacks of histological sections at cellular resolution.
These maps enable precise studies of area-specific morphological and columnar features at microscopic resolution, and in combination with existing cortical layer maps~\cite{Wagstyl2020} an investigation into layer-specific aspects of each region.
Dense maps further enable straightforward mapping from the volume to the whole brain mesh surface, which in turn facilitates comparison with other modalities, especially in-vivo imaging.
They represent an important contribution for using BigBrain as a microscopic resolution reference space, since they provide direct links to probabilistic cytoarchitectonic reference parcellations at the macroscopic scale~\cite{Amunts2020} that are widely used in neuroimaging studies.
As such, our work makes an important contribution to linking neuroscientific findings across spatial scales.

\section*{CRediT authorship contribution statement}%
\label{sec:credit_authorship_contribution_statement}


\textbf{Christian Schiffer:} Conceptualization, Methodology, Software, Validation, Formal Analysis, Investigation, Data Curation, Writing - Original Draft, Writing - Review \& Editing, Visualization.
\textbf{Hannah Spitzer:} Conceptualization, Methodology, Software, Writing - Review \& Editing.
\textbf{Kai Kiwitz:} Data Curation, Validation, Writing - Review \& Editing.
\textbf{Nina Unger:} Data Curation, Validation, Writing - Review \& Editing.
\textbf{Konrad Wagstyl:} Software, Writing - Review \& Editing.
\textbf{Alan C. Evans:} Resources, Writing - Review \& Editing, Funding acquisition.
\textbf{Stefan Harmeling:} Writing - Review \& Editing, Supervision.
\textbf{Katrin Amunts:} Conceptualization, Validation, Writing - Review \& Editing, Supervision, Project administration, Resources, Funding acquisition.
\textbf{Timo Dickscheid:} Conceptualization, Methodology, Supervision, Writing - Original Draft, Writing - Review \& Editing, Project administration, Funding acquisition.

\section*{Acknowledgments}
\label{section:acknowledgments}

This project received funding from the European Union’s Horizon 2020 Research and Innovation Programme, grant agreements 785907 (HBP SGA2) and 945539 (HBP SGA3), and from the Helmholtz Association’s Initiative and Networking Fund through the Helmholtz International BigBrain Analytics and Learning Laboratory (HIBALL) under the Helmholtz International Lab grant agreement InterLabs-0015.
This work was further supported by the German Federal Ministry of Education and Research (BMBF) and the Max Planck Society for the Advancement of Science through their joint initiative between German Universities and German Research Organizations
Computing time was granted through JARA-HPC on the supercomputer JURECA at Jülich Supercomputing Centre (JSC) as part of the project CJINM16.

\section*{Ethics Statement}
\label{section:ethics}

The study carried out requires no separate ethical approvals.
Postmortem brains were obtained in accordance to legal and ethical regulations and guidelines.
Brain tissue for datasets \bone{} and \bb{} was obtained through the body donor program of the department of anatomy of the Heinrich Heine University Düsseldorf and with approval of the ethics committee of the medical faculty of the Heinrich Heine University Düsseldorf.
Brain tissue for dataset \ba{} was obtained from the University of Maryland Brain and Tissue Bank and with approval by the Human Investigation Committees and Institutional Ethics Committees of the University of Maryland.

\section*{Declaration of competing interest}%
\label{sec:declaration_of_competing_interest}

The authors declare no competing interests.






\bibliographystyle{elsarticle-num}
\bibliography{library.bib}





\end{document}